\documentclass{aastex}
\usepackage{spr-astr-addons}
\usepackage{multirow}
\usepackage{url}
\usepackage[plainpages=false, colorlinks=true, anchorcolor=blue, linkcolor=blue, citecolor=blue, bookmarks=false]{hyperref}

\RequirePackage{color}

\begin{document}

\title{Classification of Gamma-Ray Burst durations using robust model-comparison techniques}
\slugcomment{Not to appear in Nonlearned J., 45.}
\shorttitle{Classification of GRBs}
\shortauthors{S. Kulkarni and S. Desai}

\author{Soham Kulkarni\altaffilmark{1}} 
\and  \author{Shantanu Desai\altaffilmark{2}}
\email{kulkarni.soham@gmail.com}
\email{shntn05@gmail.com}
\altaffiltext{1,2}{Department of Physics, IIT Hyderabad, Kandi, Telangana-502285, India}


\begin{abstract}
Gamma-Ray Bursts (GRBs) have been conventionally bifurcated into two distinct categories   dubbed ``short''  and ''long'', depending on whether their durations are less than or greater than two seconds respectively. However, many authors have pointed to the existence of a third  class of GRBs  with mean durations intermediate between the short and long GRBs. Here, we apply multiple model comparison techniques to verify these claims. For each category, we obtain the best-fit parameters by maximizing a likelihood function based on a weighted superposition of  two (or three) lognormal distributions. We then do model-comparison between each of these hypotheses
by comparing the chi-square probabilities, Akaike Information criterion (AIC), and  Bayesian Information criterion (BIC).  We uniformly apply these techniques to GRBs from Swift (both observer and intrinsic frame), BATSE, BeppoSAX, and Fermi-GBM.  We find that  the Swift GRB distributions (in the observer frame) for the entire dataset  favor three categories at about $2.4\sigma$ from difference in chi-squares,  and show decisive evidence in favor of  three components using both AIC and BIC. However, when the same analysis is done for  the subset of Swift GRBs with measured redshifts, two components are favored with marginal significance.  For all the other datasets, evidence for three components is either very marginal or disfavored.

\end{abstract}

\keywords {GRBs; Maximum Likelihood; Chi-Square; Bayesian Information Criterion; Model Comparison; Akaike Information Criterion}

\section{Introduction}

Gamma-ray bursts (GRBs)  are short-duration energetic cosmic explosions with prompt emission  between KeV-GeV energies, and were first detected by the Vela military
satellites  in the late 1960’s and continue to be detected at the rate of about one per day~\citep{Zhang16}. Data from the Burst and Transient
Source Explorer (BATSE) onboard the Compton Gamma
Ray Observatory (CGRO) was analyzed by ~\citet{Kouveliotou}, and led to
establishing the conventional classification of GRBs into short
(T90 $<$ 2 s) and long (T90 $>$ 2 s) classes, where T90 is the time which encompasses 90\% of the burst’s fluence, and is used as a proxy for  the duration of a GRB. 
Most classification studies of GRBs have been done using T90, although other measures have also been proposed~\citep{Zhang06,Li16}. The progenitors of long
GRBs consist of  supernovae related to the  collapse
of massive stars~\citep{Woosley} and those of  short GRBs are thought to be binary compact object mergers
~\citep{Nakar}. There are however exceptions to this general picture~\citep{Zhang,Bromberg}.

It has been observed that T90  exhibits lognormal
distributions, which were thereafter fit to short and long
GRBs~\citep{McBreen,Koshut,Kouveliotou96,Horvath2002}.
The existence of an intermediate-duration GRB class,
 with T90 in the range 2−-10s in the BATSE dataset was
first put forward by ~\citet{Horvath98,Mukherjee}. This was  
confirmed  from further analysis of the complete BATSE dataset~~\citep{Horvath2002,Chattopadhyay,Zitouni}. Evidence for a third lognormal component was also found in Swift/BAT data
~\citep{Horvath2008, Huja09, Horvath10} using duration and also from two-dimensional clustering using both duration and hardness~\citep{Veres}. This was recently corroborated for Oct 2015 Swift GRB catalog consisting of 888 GRBs by~\citet{Horvath16}, who  pointed  that three lognormal distributions provide a better fit to the data than two with 99.9999\% confidence level.~\citet{Tarnopolski16b} finds that for the similar Swift GRB dataset consisting of 947 GRBs, three groups are favored in the observer frame, whereas  two in the rest frame.  The possible physical origin of the third intermediate class of Swift GRBs is attributed to X-ray flashes~\citep{Veres}.

However, many other groups have reached opposite conclusions with the Swift and other GRB datasets. An early analysis by ~\citet{Zhang08} showed that the T90 distribution (in both the observer and rest frame)  for the first 95 Swift GRBs obeys a lognormal distribution with two components, instead of three.~\citet{Jiang}  have  found after applying  the Gaussian mixture model  on  T90 and hardness  ratio on  Swift GRBs with redshifts, two components are favored compared to three or more in both the observer and intrinsic frame. Another recent analysis of BATSE, Swift, and Fermi/GBM using application of  Gaussian Mixture model on T90 finds that BATSE, Fermi and Swift data (for GRBs with measured redshifts in observer and rest frame) are better fitted by two components~\citep{Yang}. The same analysis finds that the full Swift data (after splitting into two epochs) is consistent with three components in the observer frame. Hence, there is no uniform consensus among the  authors inspite of analyzing the same GRB datasets.

To resolve these conflicting results, we apply multiple model-comparison techniques on the distribution of T90 in a uniform manner on all GRB datasets to determine the optimum number of GRB classes. These include both frequentist hypothesis testing methods as well as  Bayesian procedures such as  Akaike Information Criterion and Bayesian Information Criterion. These model comparison techniques have been applied to a variety of problems in astrophysics and particle physics (See~\citet{Shafer,Desai16a,Desai16b} and references therein for some examples). We apply these methods to data  from multiple  detectors including BATSE, Fermi-GBM, BeppoSAX, and Swift. 

The outline of this paper is as follows. In Sect. 2 we discuss the methodology used to obtain the best-fit parameters for the mean GRB duration and its variance, after positing two and three classes of GRBs. In Sect. 3, we discuss various techniques used for model comparison. We then present results for various GRB datasets in Sect. 4, including a very brief comparison with previous results.  We conclude in Sect. 5.


\section{Parameter Estimation}

\subsection{Datasets}

Herein, we consider the GRB datasets available from BATSE\footnote{http://gammaray.msfc.nasa.gov/batse/grb/catalog/current} 4B catalog~\citep{BATSE}, Swift\footnote{http://swift.gsfc.nasa.gov/archive/grbtable}~\citep{Swift}, Fermi-GBM\footnote{http://heasarc.gsfc.nasa.gov/W3Browse/fermi/fermigbrst}~\citep{GBM} and BeppoSAX\footnote{https://heasarc.gsfc.nasa.gov/docs/sax/sax.html}~\citep{BeppoSax}. The number of GRBs analyzed for the model comparison are 2036 from BATSE, 927 from Swift, 1901 from Fermi, and 1003 from BeppoSAX. These detectors account for almost all the GRBs discovered in the past three decades.  We did not consider other catalogs such as those from RHESSI, INTEGRAL etc, as they contained less than 500 GRBs. 

\subsection{Fitting method}

We have applied the same Maximum Likelihood (ML) method as proposed in ~\citet{Horvath16} (see also ~\citet{Horvath2002,Horvath2008,Horvath09}) for fitting the data and obtain the best-fit parameters. As is done in a ML method, we select a probability density function and define a log-likelihood function, which is to be maximized over the parameter space by varying the free parameters. We model the probability density function to be a superposition of $k$ lognormal gaussian  distributions, where $k$ is the total number of GRB classes. Also, we associate a weight $w_{j}$ for each $k$, which indicates the number of GRBs of that particular type found in our dataset. For a  probability density function $f(x,\theta) $, where $\theta $ is the set of parameters required for defining the probability function, the log-likelihood will be defined as:
\begin{equation}
\mathcal{L} = \sum\limits_{i=1}^{N} \ln \sum_{j=1}^{k} w_{j}f_{j}(x_{i},\theta), 
\label{eq:likelihood}
\end{equation}  

\noindent where $x_{i}$ are the sample datapoints (in our case log of the T90 distribution), $w_{j}$ is the number of GRB categories, and $N$ is the total number of GRBs analyzed. As stated above we take the $k$ lognormal distributions of the form:
\begin{equation}
f(x,\theta) = \frac{1}{\sqrt{2\pi} \sigma} \exp\left(- \frac{(x-\overline{\log T_{90}})^{2}}{2\sigma^{2}}\right),
\end{equation}
\noindent where $\overline{\log T_{90}}$ is mean of the logarithm of T90 distribution for each class,  and the weights $w_i$
satisfy the condition:
\begin{equation}
\sum\limits^{k}_{i=1} w_{i} = N.
\label{eq:weights}
\end{equation}

The maximization is done by implementing an  optimization algorithm ({\tt SLSQP} and {\tt COBYLA}) included in the {\tt SciPy} package of Python. We did not bin the data during  the optimization process. We should also point out that the optimization of the likelihood in Eq.~\ref{eq:likelihood}  is mathematically similar to  Gaussian Mixture Model (GMM), if the weights in Eq.~\ref{eq:weights} are normalized to unity and the covariance matrix is diagonal~\citep{astroml}. The parameters of the GMM can be found by
the Expectation-Maximization (EM) Algorithm~\citep{astroml}.  The GMM and the corresponding parameter estimation using the EM algorithm have been also applied to GRB datasets using both T90~\citep{Yang} as well as using T90 vs hardness ratio~\citep{Jiang}.  Note however that ~\citet{Yang} have included the covariances between the datasets.  We also tried
to estimate the best-fit parameters by applying the EM algorithm after normalizing the weights to unity, instead of the total number of GRBs and assuming covariances are diagonal. However, the best likelihood model is still obtained by using the optimization algorithm in {\tt SciPy} and in the rest of the paper, we report the best-fit values from this.

\section{Model Comparison}
\label{sec:modelcomparison}
The comparison of models on the basis of best-fit likelihood (or minimum $\chi^2$) is not a good way to  do hypothesis testing or select the optimum model after finding the best-fit parameters for each model. As we increase the number of free parameters, it is obvious that the likelihood will increase, but it leads  to over-fitting. Therefore, the additional free parameters need to be  penalized so as to avoid getting a bad result. This is called \emph{Occam's Razor}.  To address these issues,
a number of both frequentist and Bayesian model-comparison techniques have been used
over the past decade to determine the best model which fits the observational data~\citep{Liddle,Liddle07,Liddle2,Lyons}. Here, we use multiple analysis methods, such as the frequentist hypothesis testing (based on $\chi^2$ probabilities) and information criterion based tests such as  Akaike Information Criterion (AIC) and  Bayesian Information Criterion (BIC) 
for  model comparison. AIC and BIC have also been previously used  for GRB classification by a number of authors~\citep{Mukherjee,Tarnopolski16b,Tarnopolski,Jiang,Yang}.
Frequentist model comparison after binning the data has been used by ~\citet{Zitouni,Tarnopolski15}.
More information about AIC and BIC and its application to a variety of astrophysical problems can be found in ~\citet{Liddle,Liddle07,Biswas,Shi,Shafer}.  We should point that these are not the only possibilities for model comparison. Other techniques include Bayes factor, posterior odds, $p$ values of test statistics, etc. These are extensively discussed in ~\citet{Liddle, Liddle2, Lyons} and references therein. However, some of them are computationally very intensive. Unfortunately, there  is no golden rule to decide which among the above methods is best suited for a given problem. For any model comparison problem, it is therefore important to apply multiple methods and test that they lead to consistent results. Unfortunately, there is also no simple answer as to what to do in case multiple  methods used for model comparison provide conflicting results, except for validations with Monte-Carlo simulations or using mock catalogs. We now discuss the three methods used for model comparison in this work.

\subsubsection{Chi-Square Test}
In order to construct a frequentist model comparison test, we calculate the reduced $\chi^2$ to compare different models.   The reduced $\chi^2$ is equal  to $\chi^2/\nu$, where $\nu$ is the total degrees of freedom and   $\chi^2$ is defined as follows:

\begin{equation}
\chi^{2} = - \sum\limits^{N}_{i=1} \ln \sum\limits_{j=1}^{k} \frac{w_{j}}{\sqrt{2\pi} \sigma_j} \exp \left( \frac{-(x_{j}-\overline{\log T_{90}})^{2}}{2\sigma_{j}^{2}}\right);
\label{eq:chi}
\end{equation} 
under the condition that
\begin{equation}
\sum\limits^{k}_{i=1} w_{i} = 1
\label{eq:wt}
\end{equation}

This is essentially the same expression (modulo the minus sign) that we are using for the calculation of the likelihood in Eq.~\ref{eq:likelihood}, except that the weights in Eq.~\ref{eq:wt} are now normalized to unity instead of the total number of GRBs.  We have also verified using numerical simulations that for a distribution of two Gaussians,   $\chi^2/\nu \sim 1$ for the best-fit input parameters, where
 $\chi^2$ is defined in Eq.~\ref{eq:chi}. We note that while constructing this $\chi^2$, we have not binned the data in T90.  After obtaining  the best-fit model parameters for each hypothesis, we compare the  $\chi^2$ probability, after  taking into account the total degrees of freedom. The  $\chi^2$ probability is equal to $\frac{1}{2^{\nu/2}\Gamma(\nu/2)}(\chi^2)^{\nu/2-1}\exp(-\chi^2/2)$~\citep{NR92}, where $\Gamma$ is the incomplete Gamma function and $\nu$ is the total degrees of freedom. 
The preferred model is the one with the higher value of $\chi^2$ probability. If two models
are nested, then according to Wilk's theorem~\citep{Wilks}, the difference in $\chi^2$  between the
two models satisfies a $\chi^2$ distribution with degrees of freedom equal to  the 
 difference in the number of free parameters for the two hypotheses~\citep{Lyons}. Since a model with two Gaussian components is a special case of a model with three components, we can apply Wilk's theorem to assess the statistical significance of the better model.

\subsubsection{AIC}

The Akaike Information Criterion (AIC) is used for model comparison, when we need to penalize for any additional free parameters  to avoid overfitting. A preferred model in this test is the one with the smaller value of AIC between the two hypothesis. AIC is an approximate  minimization of Kullback–-Leibler information entropy, which estimates the distance between two probability distributions~\citep{Liddle07}. The AIC is given by:

\begin{equation}
AIC = 2p - 2 \ln L
\label{eq:aic}
\end{equation}

\noindent where $p$ is the number of free parameters in the model and $L$ is the likelihood.
The AIC defined in Eq.~\ref{eq:aic} is good when the ratio $N/p$ is very large i.e. $>$ 40~\citep{Burnham}. For
 a smaller value of the ratio, a first order correction is included and the expression is:

\begin{equation}
AIC = 2p - 2 \ln L + \frac{2p(p+1)}{N-p-1}
\end{equation}

As all our datasets have a ratio of $N/p$ greater than 40, we don't need to worry about this correction. The absolute value of AIC is usually not of interest. The goodness of fit between two hypothesis (A) and (B) is described by the difference of the AIC values and is given by:

\begin{equation}
\Delta AIC = AIC_{A} - AIC_{B},
\label{eq:AIC}
\end{equation}
where  $AIC_{A}$ - $AIC_{B}$ correspond to the AIC values for the  hypothesis A and B. \citet{Burnham} have provided qualitative
strength of evidence rules to assess the significance of a model based on the $\Delta$AIC  values between the two models. If $\Delta$AIC$>5$, then it is considered strong evidence against the model with higher AIC and 
  $\Delta$AIC$>10$ is considered as decisive evidence against the model with higher AIC~\citep{Liddle07}. Values of $\Delta$AIC$<5$ correspond to weak evidence.

\subsubsection{BIC}

The Bayesian Inference Criterion (BIC) is also used for penalizing the use of extra parameters. BIC is an approximation for Bayesian evidence. As in the case of AIC, the model with the smaller value of BIC is the preferred model. The penalty in the BIC test is harsher than that in the case of AIC and is given by:

\begin{equation}
BIC = p \ln N - 2 \ln L
\end{equation} 

The logarithmic term and the number of free parameters act as a very harsh measure needed for the BIC test. The goodness of fit used
for hypothesis testing between two models $A$ and $B$ is given by:
\begin{equation}
\Delta BIC = BIC_{A} - BIC_{B}
\label{eq:BIC}
\end{equation}
Similar to AIC, the model with lower value of BIC is favored. To assess the significance of a model, strength of evidence rules have also been proposed based on $\Delta$BIC~\citep{Robert}, which are approximately the same as those for AIC. 
We note that other information theoretic criterion have also been proposed besides AIC and BIC and these are discussed in ~\citet{Liddle07}.

\section{Results}
We apply all the techniques discussed in the previous section to GRB datasets from various detectors. For data from each of the GRB detectors, we find the mean value of  T90 and its standard deviation by positing that the data has two as well as three components, followed by  maximizing the likelihood in Eq.~\ref{eq:likelihood} for both the hypotheses. For these best-fit parameters, we then implement all the three  model-comparison techniques outlined in Sect.~\ref{sec:modelcomparison}. For the calculation of information criterion,  since we are comparing the two-component model vs the three-component, we consider the two-component GRB as the null hypothesis and calculate $\Delta$AIC and $\Delta$BIC (cf. Eq.~\ref{eq:AIC} and ~\ref{eq:BIC}) as the difference between the AIC/BIC value for the three-component fit and the two-component fit. Therefore, if $\Delta$AIC/BIC$>0$, then the two-component model  is favored and vice-versa.  We now present our results for BATSE, BeppoSAX, Fermi-GBM and Swift.

\begin{figure}
\includegraphics[width=8cm]{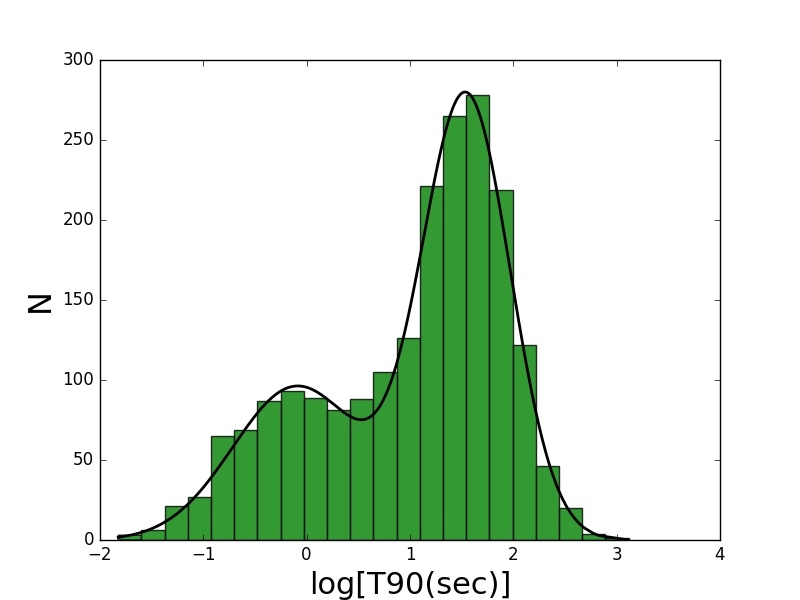}
\caption{A fit for the 2-component model for BATSE GRBs. Details of the fits can be found in Table~\ref{tab:batse}.}
\label{fig:batse1}
\end{figure}

\begin{figure}
\includegraphics[width=8cm]{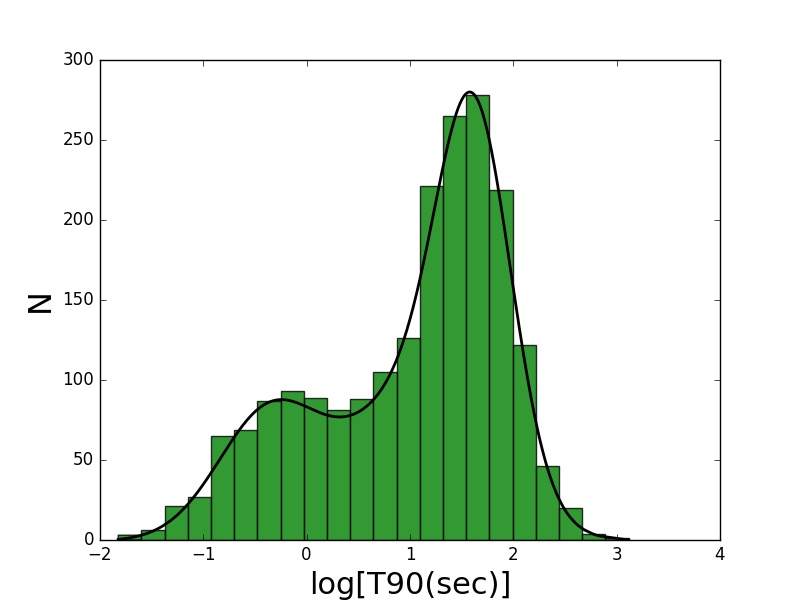}
\caption{A fit for the 3-component model for BATSE GRBs. Details of the fits can be found in Table~\ref{tab:batse}.}
\label{fig:batse2}
\end{figure}

\begin{table*}[!htbp]
\caption{Model Comparison Parameters for BATSE GRBs. The first column contains the total number of GRB classes and the next three indicate the best-fit values for the logarithm of the mean T90 ($\mu_{T90}$), its standard deviation ($\sigma_{T90}$), total number of GRBs ($w_i$) in each category after positing both two and three types of GRBs. These are obtained by maximizing Eq.~\ref{eq:likelihood}. $\mathcal{L}$, $p(\chi^2,\nu)$, $\chi^2/dof$,  AIC ,  BIC  represent the likelihood, $\chi^2$ probability for $\nu$ DOF, reduced $\chi^2$, Akaike and Bayesian Information criterion respectively. The last three columns indicate the $p$-value, $\Delta$AIC, and $\Delta$BIC between the three component and two-component model, which are used for model comparison. We have not used the likelihoods for model comparison.  In the table the preferred value for every test is highlighted in bold. We note that if $\Delta$AIC or $\Delta$BIC$>0$, then two GRB classes are preferred and vice-versa. We find that AIC and $\chi^2$ probability favor three components, whereas BIC favors two. However with all these model comparison techniques, the statistical significance is marginal.}
\label{tab:batse}
\begin{tabular}{|c|ccc|ccccc|ccc|}
\hline
$k$ & $\mu_{T90}$ & $\sigma_{T90}$ & $w_{i}$ & $\mathcal{L}$ & $\chi^{2}/\nu$   & P($\chi^{2},\nu$) &  AIC  &  BIC  & $p$-value & $\Delta(AIC) $& $ \Delta(BIC)$ \\
\hline
\multirow{2}{*}{2} & -0.093 & 0.62 & 681 & \multirow{2}{*}{13076} & \multirow{2}{*}{1.199}  & \multirow{2}{*}{1.04e-10}    & \multirow{2}{*}{4879.5} & \multirow{2}{*}{\textbf{4902}}& \multirow{5}{*}{0.108(1.2$\sigma$)} & \multirow{5}{*}{-6.5} & \multirow{5}{*}{5} \\
\cline{2-4}
& 1.542 & 0.43 & 1355 & & &  & & & & &\\
\cline{1-9}
\multirow{3}{*}{3} & -0.361 & 0.5 & 459 & \multirow{3}{*}{13081} & \multirow{3}{*}{1.197}  & \multirow{3}{*}{\textbf{1.3e-10}}  & \multirow{3}{*}{\textbf{4873}} & \multirow{3}{*}{4907} & & & \\
\cline{2-4}
& 1.09 & 0.65 & 692 & & & &  & & & &\\
\cline{2-4}
& 1.63 & 0.36 & 885 & & & &  & & & &\\
\hline
\end{tabular}
\end{table*}

\subsection{BATSE}

The current BATSE  GRB~\citep{BATSE} catalogue contains 2036 GRBs detected between 1991 and 2000. The fits for the data for $k=2$ and $k=3$  are shown graphically in Figs.~\ref{fig:batse1} and ~\ref{fig:batse2} respectively. From the figures, we can see that both the fits are indistinguishable by eye.  When we fit the BATSE dataset for three components, we find that 459 GRBs belong to the short, 692 to the intermediate, and 885 to the long category. While fitting for two components, we find that
681 GRBs belong to the short category and 1355 to the long category. The detailed results of model comparison are tabulated in Table no.~\ref{tab:batse}.  This table contains the likelihood, reduced $\chi^2$, AIC, and BIC for both the hypotheses. Here, we find that $\Delta$AIC = -6.5. This corresponds to  strong evidence for  three components. When we compare the $\chi^2$ probability, we find that $k=3$ has a higher value, which implies that it is a better fit.  To assess the statistical significance of $k=3$ model compared to $k=2$  model, we apply Wilk's theorem  and find that the $p$-value is equal to  0.108. This implies  that there is 10.8\% probability that the third component is a statistical fluctuation. This $p$-value  corresponds to $1.2\sigma$ Gaussian significance~\citep{NR92}.  However, the $k=2$ model has a lower value of BIC and $\Delta$BIC between the two models is equal to 5, corresponding to weak evidence.

Therefore  in summary, two of the three model comparison techniques (AIC and frequentist test) prefer $k=3$ and one of them (based on BIC) prefer $k=2$. However, in all the three cases, the significance of one model with respect to the other is marginal and none of these tests pass the $5\sigma$ criterion (usually used in high-energy physics) to decisively pick one model over the other. We note that from similar likelihood analysis of the BATSE data   and comparison of likelihoods,~\citet{Horvath2002} found evidence for three GRB classes  and the probability that the third group is a fluctuation is 0.5\%. On the other hand, a recent GMM-based analysis of BATSE T90 dataset showed evidence for  two components with $\Delta$BIC=13 in favor of two component model~\citep{Yang}. One possible reason for their higher value of $\Delta$BIC=13  compared to ours, could be that in ~\citet{Yang}, the covariances between different data points have been taken into account. However, we should emphasize that from our analysis the significance of the third component from AIC and frequentist test is marginal.

\begin{figure}
\includegraphics[width=8cm]{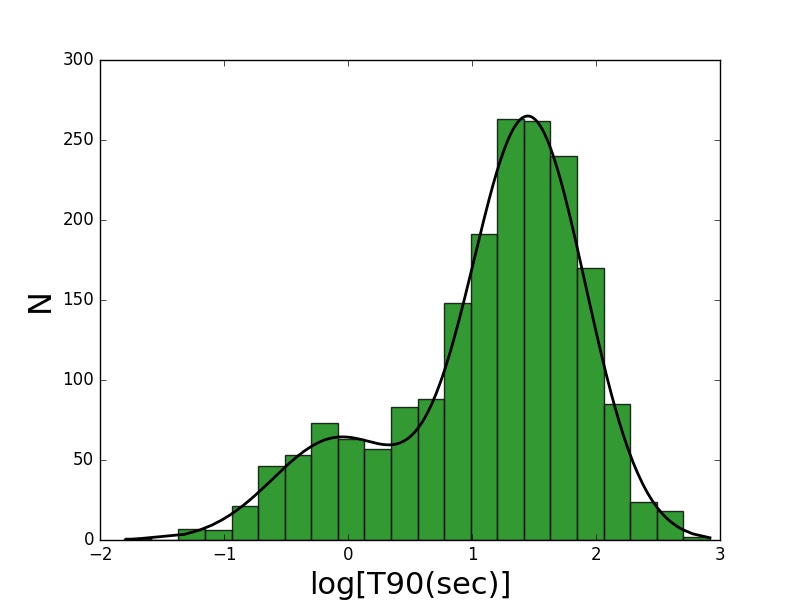}
\caption{A fit for the 2-component model for FERMI GBM GRBs. Summary of the fits can be found in Table~\ref{tab:fermi}.}
\label{fig:fermi1}
\end{figure}

\begin{figure}
\includegraphics[width=8cm]{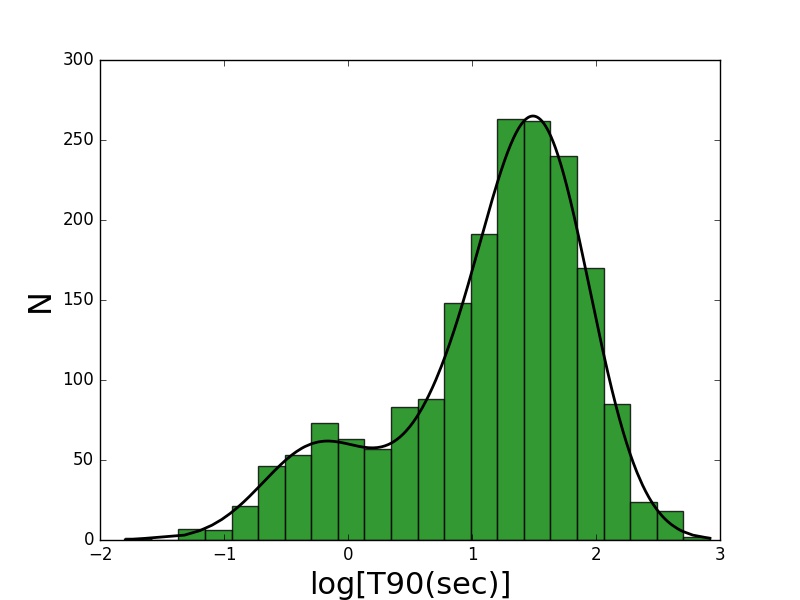}
\caption{A fit for the 3-component model for FERMI GBM GRBs. Summary of the fits can be found in Table~\ref{tab:fermi}.}
\label{fig:fermi2}
\end{figure}

\begin{table*}[!htbp]
\caption{Model Comparison Parameters Fermi-GBM GRBs. The explanation of all columns can be found in Table~\ref{tab:batse}. In the table the preferred value for every test is highlighted in bold. We find that AIC and $\chi^2$ probability favor three components, whereas BIC favors two. However with all these model comparison techniques, the statistical significance is marginal.}
\label{tab:fermi}
\begin{tabular}{|c|ccc|ccccc|ccc|}
\hline
$k$ & $ \mu_{T90}$ & $\sigma_{T90} $ & $w_{i}$ & $\mathcal{L}$   &$\chi^{2}/\nu$ &P($\chi^{2},\nu$) &  AIC  &  BIC  & $p$-value & $\Delta(AIC) $& $ \Delta(BIC)$ \\
\hline
\multirow{2}{*}{2} & -0.0851 & 0.52 & 406 & \multirow{2}{*}{12314} & \multirow{2}{*}{1.075} & \multirow{2}{*}{4.74e-03}  & \multirow{2}{*}{4086.7} & \multirow{2}{*}{\textbf{4109}}&\multirow{5}{*}{0.32(0.5$\sigma$)} &\multirow{5}{*}{-1.7} & \multirow{5}{*}{10}\\
\cline{2-4}
& 1.45 & 0.46 & 1495 & & & & &  & & & \\
\cline{1-9}
\multirow{3}{*}{3} & -0.251 & 0.443 & 309 & \multirow{3}{*}{12316} & \multirow{3}{*}{\textbf{1.0748}} & \multirow{3}{*}{\textbf{4.808e-03}}  & \multirow{3}{*}{\textbf{4085}} & \multirow{3}{*}{4119} & & & \\
\cline{2-4}
& 1.133 & 0.531 & 691 & & & & &  & & &\\
\cline{2-4}
& 1.589 & 0.402 & 901 & & & & &  & & &\\
\hline
\end{tabular}
\end{table*}

\subsection{Fermi-GBM}

The Fermi-GBM catalogue (as of Sept. 2016) currently has 1901 GRBs ~\citep{GBM}. 
When we fit the data for three components we find that 309 are short, 691 are intermediate, and 901 belong to long type. On positing two components, we find that 406 GRBs belong to the short category and 1495 belong to the long category.  The model fits to the data shown  in Figs.~\ref{fig:fermi1} and \ref{fig:fermi2} are not much different  compared to those for  BATSE GRBs. Similar to BATSE, both AIC and the frequentist comparison test favor the $k=3$ case over the $k=2$ case, whereas BIC prefers two components. 
The model comparison values from Table~\ref{tab:fermi} are $\Delta$AIC = $-1.7$,  which favor the $k=3$ model very weakly. However,  $\Delta$BIC = 10, which corresponds to strong evidence for $k=2$ model.
The frequentist test using $\chi^{2}$ probability prefers three components with  $p$-value = 0.3, which  only corresponds to $0.5\sigma$ significance.

Therefore,  to summarize, AIC and frequentist model comparison for Fermi-GBM GRBs prefer three components, whereas BIC prefers two. However, the statistical significance of all the three tests is marginal,  thus implying that both hypothesis cannot be easily distinguished. Our results also agree with the  analysis of ~\citet{Tarnopolski15}, who also compared $\chi^2$ probabilities by carrying out a binned analysis of the duration distribution. He concluded that although the $\chi^2$ probability for a 3-component fit is more than a 2-component one, the $p$-values range from 14-77\% for different values of the binning. Therefore from this analysis, there is no evidence that the third peak is statistically significant. The GMM based analysis by ~\citet{Yang} using 1741 GRBs shows a preference for two components with $\Delta$BIC=13, which approximately agrees with our value of $\Delta$BIC, although our value does not cross the decisive evidence threshold of greater than 10. The significance of the third component from our analysis using  the other two tests is very marginal.

\begin{figure}
\includegraphics[width=8cm]{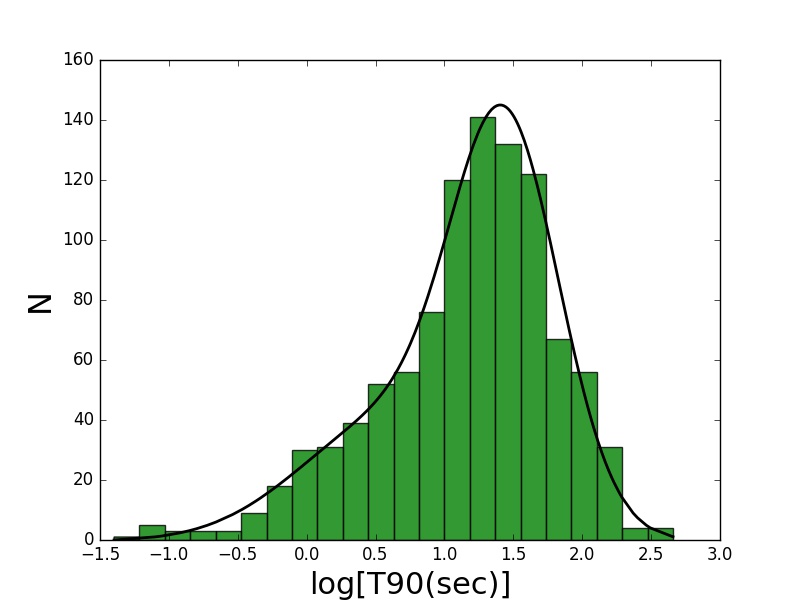}
\caption{A fit for the 2-component model for BeppoSAX GRBs. Summary of the fits can be found in Table~\ref{tab:beppo}.}
\label{fig:beppo1}
\end{figure}
\begin{figure}
\includegraphics[width=8cm]{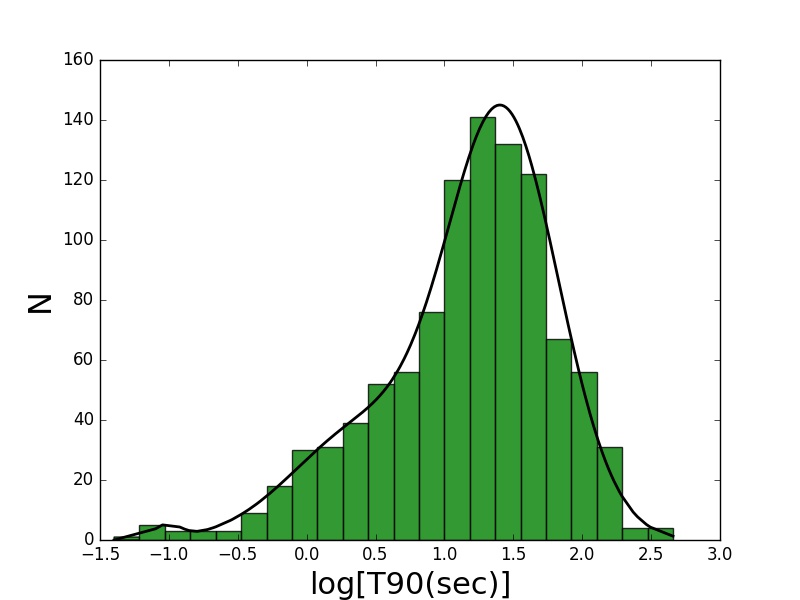}
\caption{A fit for the 3-component model for BeppoSAX GRBs. Summary of the fits can be found in Table~\ref{tab:beppo}.}
\label{fig:beppo2}
\end{figure}

\begin{table*}[!htbp]
\caption{Model Comparison Parameters for BeppoSAX GRBs. Explanation of all columns can be found in Table~\ref{tab:batse}. The preferred value for every test is highlighted in bold. We find that BIC and $\chi^2$ probability favor two components, whereas AIC favors three. However with all these model comparison  techniques, the statistical significance is marginal.}
\label{tab:beppo}
\begin{tabular}{|c|ccc|ccccc|ccc|}
\hline
$k$ & $ \mu_{T90}$ & $\sigma_{T90} $ & $w_{i}$ & $\mathcal{L}$ &  $\chi^{2}/\nu$ &$P(\chi^{2},\nu)$ &  AIC  &  BIC  & $p$-value & $\Delta(AIC) $& $ \Delta(BIC)$ \\
\hline
\multirow{2}{*}{2} & 0.626 & 0.668 & 356 & \multirow{2}{*}{6012} & \multirow{2}{*}{0.921} & \multirow{2}{*}{\textbf{0.00185}}  & \multirow{2}{*}{1847} & \multirow{2}{*}{\textbf{1867}} &\multirow{5}{*}{0.256(0.6$\sigma$)} & \multirow{5}{*}{-1.5}& \multirow{5}{*}{8}\\
\cline{2-4}
& 1.45 & 0.392 & 647 & & & &  & & & &\\
\cline{1-9}
\multirow{3}{*}{3} & -1.013 & 0.1 & 6 & \multirow{3}{*}{6015} & \multirow{3}{*}{0.919}  & \multirow{3}{*}{0.00177} & \multirow{3}{*}{\textbf{1845.5}}   & \multirow{3}{*}{1875} & & & \\
\cline{2-4}
& 0.4307 & 0.530 & 259 & & &  & & & & &\\
\cline{2-4}
& 1.43 & 0.404 & 738 & & & &  & & & &\\
\hline
\end{tabular}
\end{table*}
\subsection{BeppoSAX} 

The BeppoSAX catalogue~\citep{BeppoSax} has a total of 1003 GRBs detected between 1996
 and 2001. The results of our fits for two and three components are shown in  Figs.~\ref{fig:beppo1} and Figs.~\ref{fig:beppo2} respectively. A tabular summary of  our model comparison tests can be found  in Table 3. 
After doing a three-component fit, we find that the GRBs from BeppoSAX are mainly divided  into intermediate and long GRB segments (259 and 738 respectively) leaving only 6 GRBs in the short category. For a two-component fit, we find that 356 GRBs belong to short category and 647 to long.
  From the data presented in Table no.~\ref{tab:beppo}, we do not have a strong consensus to decide the preferred model. Both the $\chi^2$ probability and BIC prefer $k=2$, whereas AIC shows a preference for $k=3$. We find that  $\Delta$AIC = -1.5  corresponding to weak evidence. The $\chi^{2}$ probability gives a  $p$-value of 0.256, corresponding to only $0.6\sigma$ significance.  The $\Delta$BIC = 8 in favor of the $k=2$ model corresponding to strong evidence.

Therefore, none of the three model comparison tests provide a decisive evidence for the three-component model over the two-component one or vice-versa. A likelihood analysis of the BeppoSAX data by ~\citet{Horvath09}  showed evidence for three components, but the probability that this is a fluctuation was only 3.7\%.

\begin{figure}
\includegraphics[width=8cm]{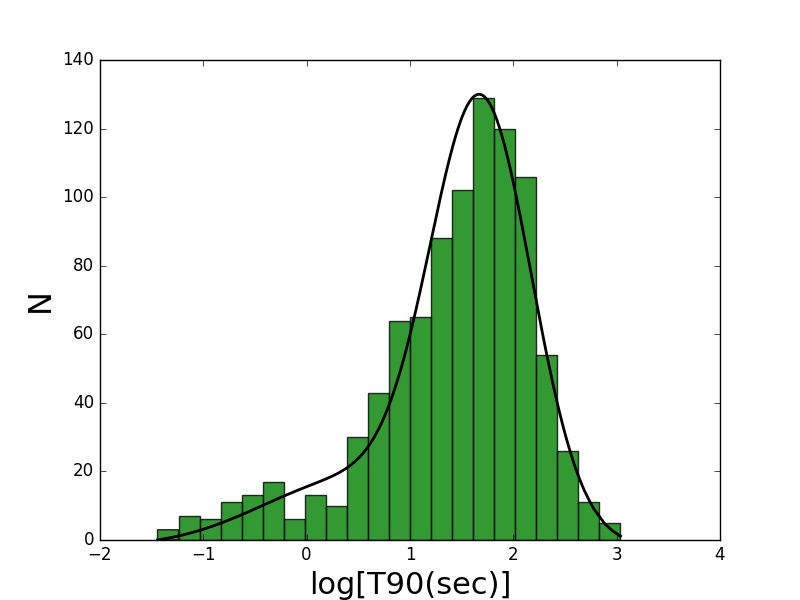}
\caption{A fit for the 2-component model for Swift GRBs. Summary of the fits can be found in Table~\ref{tab:swift}.}
\label{fig:swift1}
\end{figure}
\begin{figure}
\includegraphics[width=8cm]{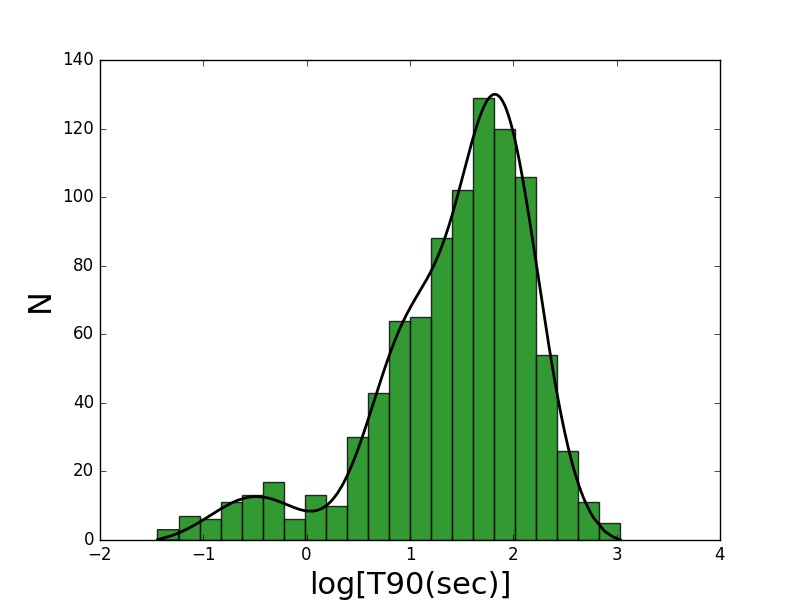}
\caption{A fit for the 3-component model for Swift GRBs.Summary of the fits can be found in Table~\ref{tab:swift}. }
\label{fig:swift2}
\end{figure}

\begin{table*}[!htbp]
\caption{Model comparison parameters for Swift GRBs. Explanation of all the columns can be found in Table~\ref{tab:batse}. The preferred value for every test is highlighted in bold. All the three tests favor three components compared to two. Both BIC and AIC point to decisive evidence (in terms of significance) for the three components and the significance of three component model compared to two is $2.36\sigma$. }
\label{tab:swift}
\begin{tabular}{|c|ccc|ccccc|ccc|}
\hline
$k$ & $ \mu_{T90}$ & $\sigma_{T90} $ & $w_{i}$ & $\mathcal{L}$ & $\chi^{2}/\nu$ &$p(\chi^{2},\nu) $  &  AIC  &  BIC  & $p-value$ & $\Delta(AIC) $& $ \Delta(BIC)$ \\
\hline
\multirow{2}{*}{2} & 0.422 & 0.869 & 200 & \multirow{2}{*}{5354} &  \multirow{2}{*}{1.075} & \multirow{2}{*}{0.002}   & \multirow{2}{*}{1997} & \multirow{2}{*}{2016} & \multirow{5}{*}{0.009(2.36$\sigma$)}&\multirow{5}{*}{-28} & \multirow{5}{*}{-18}\\
\cline{2-4}
& 1.388 & 0.428 & 727 & & & & & &  & &\\
\cline{1-9}
\multirow{3}{*}{3} & -0.492 & 0.441 & 74 & \multirow{3}{*}{5372} & \multirow{3}{*}{1.061} & \multirow{3}{*}{\textbf{0.00388}}  & \multirow{3}{*}{\textbf{1969}} & \multirow{3}{*}{\textbf{1998}} & & & \\
\cline{2-4}
& 0.98 & 0.385 & 266 & & & & & &  & &\\
\cline{2-4}
& 1.857 & 0.388 & 587 & & & & &  & & &\\
\hline
\end{tabular}
\end{table*}
\subsection{Swift GRBs}

We analyzed  927 Swift (BAT) GRBs detected between Nov. 2004 and Sept. 2016~\citep{Swift}.   The results from our likelihood fits for $k=2$ and $k=3$ are shown in Figs.~\ref{fig:swift1} and ~\ref{fig:swift2} respectively. We see from Figs.~\ref{fig:swift1} and ~\ref{fig:swift2} that the $k=3$ model gets preferred over the $k=2$ model but only slightly.  After fitting for two components, we find that 200 and 727 belong to the short and long category respectively. On doing the same for three components, we find that 74, 266, and 587 belong to short, intermediate, and long class respectively. From Table~\ref{tab:swift}, we find that the Swift catalogue prefers the $k=3$ case over $k=2$ case with all the  three tests. 
We find that both $\Delta$BIC=$-18$ and $\Delta$AIC=$-28$ have absolute values greater than 10, which corresponds to decisive evidence for a 3-component model compared a 2-component one. From our frequentist model comparison test, the $\chi^2$ probability is higher for the three-component model with a $p$-value of 0.009 corresponding to $2.36\sigma$ significance. More data is necessary to see if the significance enhances with increased data sample.

Therefore, both the information criterion based  model comparison tests point to  decisive evidence for a trifurcated GRB data sample (based on its duration). The significance of the third component from the frequentist model comparison test is  $2.36\sigma$. As mentioned in the introduction, a large number of  groups have analyzed the Swift data over the past decade. Using a maximum likelihood analysis~\citet{Horvath16} finds that three distributions fit the data better than two with 99.9999\% ($4.75\sigma$) significance. On comparing maximum likelihood, AIC, and BIC, ~\citet{Tarnopolski} also finds three distributions fit the data better than two. The results from GMM based analysis are also consistent with three distributions with $\Delta$BIC of about 6~\citep{Yang}. Note however that in ~\citet{Yang}, they have done two separate analysis of the Swift data, after bifurcating the GRB sample depending on whether they were detected before or after Dec 2012. Therefore, our results qualitatively agree with recent T90 based classifications  of the Swift dataset by other authors.

\subsection{Swift GRBs in rest frame}

\begin{figure}
\includegraphics[width=8cm]{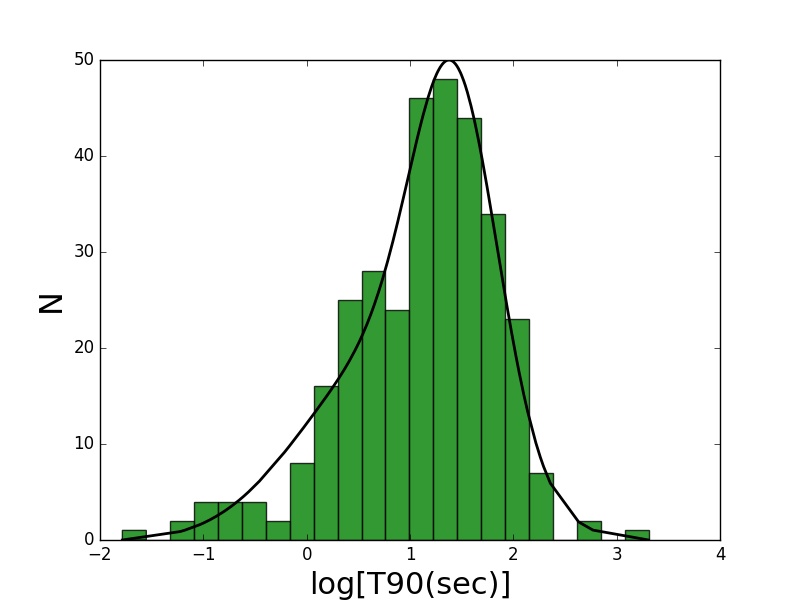}
\caption{A fit for the 2-component model for intrinsic red shifted Swift GRBs. Summary of the fits can be found in Table~\ref{tab:redss}.}
\label{fig:redss1}
\end{figure}

\begin{figure}
\includegraphics[width=8cm]{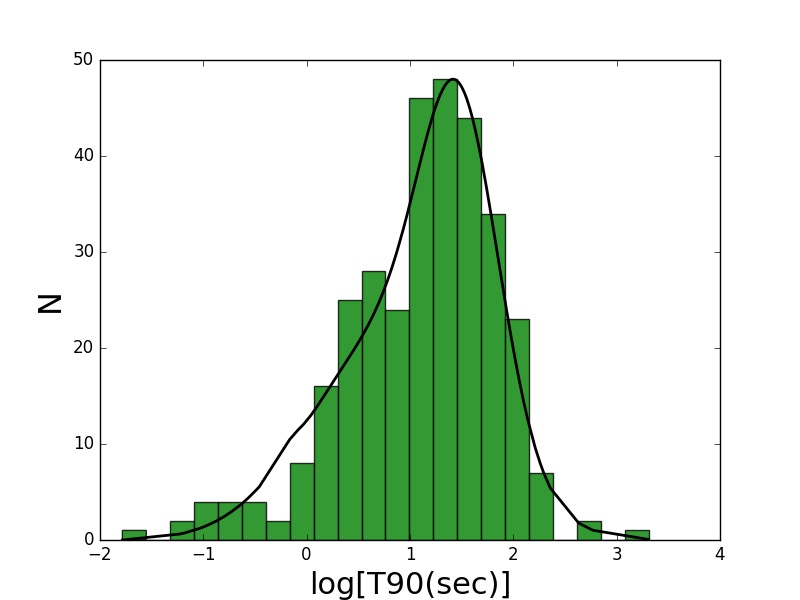}
\caption{A fit for the 3-component model for intrinsic red shifted Swift GRBs. Summary of the fits can be found in Table~\ref{tab:redss}.}
\label{fig:redss2}
\end{figure}

We now carry out a similar classification of the intrinsic durations of the GRBs by taking into account the measured redshifts. The Swift GRB catalog consists of 323 detections with redshifts.
All the other detectors have less than 100 GRBs with measured redshifts. So we restrict this analysis
to the intrinsic T90 distribution of only the Swift GRB sample.
The intrinsic durations  for a  GRB  is given by:
\begin{equation}
T_{90_{int}} = \dfrac{T_{90_{obs}}}{1+z}
\end{equation}
\noindent where as indicated, $T_{90_{obs}} $ are the measured T90 values, $T_{90_{int}}$ the intrinsic T90 values, and $z$ being the redshift for the GRB. The model fits for the intrinsic T90 for the Swift  GRBs are shown in Fig.~\ref{fig:redss1} and Fig.~\ref{fig:redss2}  respectively for the $k=2$ and $k=3$ case and a tabular summary of the model comparison results in Tab.~\ref{tab:redss}. On fitting the data to two components, we find that 163 and 160 GRBs belong to the short and long category respectively. On doing the same for three components, 1, 185, and 137 fit in short, intermediate, and long categories respectively.


\begin{table*}[!htbp]
\caption{Model comparison parameters for the intrinsic durations detected by Swift GRBs after incorporating the measured redshifts. Explanation of all the columns is same as in Table~\ref{tab:batse}. The preferred values are highlighted in bold. All tests prefer two components.  However the significance is marginal in all the cases.}
\label{tab:redss}
\begin{tabular}{|c|ccc|ccccc|ccc|}
\hline
$k$ & $ \mu_{T90}$ & $\sigma_{T90} $ & $w_{i}$ & $\mathcal{L}$ & $\chi^{2}/\nu$ &P$(\chi^{2},\nu)$   &  AIC  &  BIC  & $p$-value & $\Delta(AIC) $& $ \Delta(BIC)$ \\
\hline
\multirow{2}{*}{2} & 0.761 & 0.826 & 163 & \multirow{2}{*}{1521} & \multirow{2}{*}{1.082} & \multirow{2}{*}{\textbf{8.75e-3}} & \multirow{2}{*}{\textbf{698.5}} & \multirow{2}{*}{\textbf{713}} & \multirow{5}{*}{0.518($<0.02\sigma$)}&\multirow{5}{*}{1.7} & \multirow{5}{*}{9.8}\\
\cline{2-4}
& 1.439 & 0.4163 & 160 & & & & & & &  &\\
\cline{1-9}
\multirow{3}{*}{3} & -1.783 & 0.1 & 1 & \multirow{3}{*}{1522} & \multirow{3}{*}{1.085} & \multirow{3}{*}{8.49e-3}   & \multirow{3}{*}{700.2} & \multirow{3}{*}{722.8} & & & \\
\cline{2-4}
& 0.8239 & 0.7865 & 185 & & & & & & &  &\\
\cline{2-4}
& 1.485 & 0.3798 & 137 & & & & & & &  &\\
\hline
\end{tabular}
\end{table*}

The intrinsic T90 distribution for the Swift GRBs weakly prefers the $k=2$ case over $k=3$ case with all the three model-comparison tests. The value of $\Delta$AIC is equal to  1.7, which amounts to weak evidence. Similarly, $\Delta$BIC = 9.8,  which corresponds to strong evidence.  The frequentist model comparison test also shows a   preference for $k=3$ model with a $p$-value of $0.518$ corresponding to less than $0.02\sigma$. In summary, we can say that the preferred model at first look is the $k=2$ model for the intrinsic GRB case but its significance is low with all the model comparison tests used.   When a similar analysis of the intrinsic T90 distribution of 347 Swift GRBs  was done by ~\citet{Tarnopolski}, he found that  AIC  points to three components (albeit with very weak evidence against two), and BIC yielded a very strong support for two components.  The GMM-based analysis also showed evidence for two components with $\Delta$BIC of about 6~\citep{Yang}. Therefore, our results qualitatively agree with similar analysis by other authors.

\subsection{Swift GRBs with measured redshifts in observer frame}

\begin{figure}
\includegraphics[width=8cm]{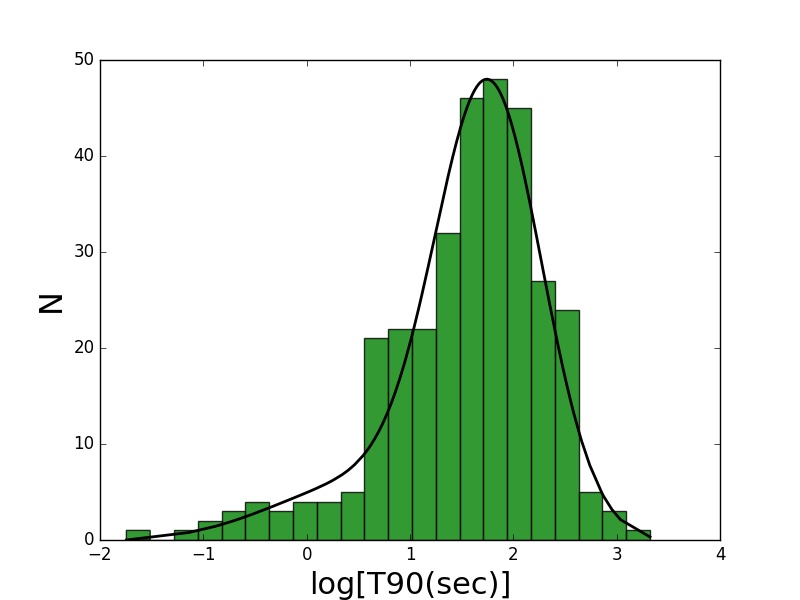}
\caption{A fit for the 2-component model for a subset of SWIFT GRBs with measured redshifts the observer frame. 
Summary of the fits can be found in Table~\ref{tab:redss_new}.}
\label{fig:redss_new1}
\end{figure}

\begin{figure}
\includegraphics[width=8cm]{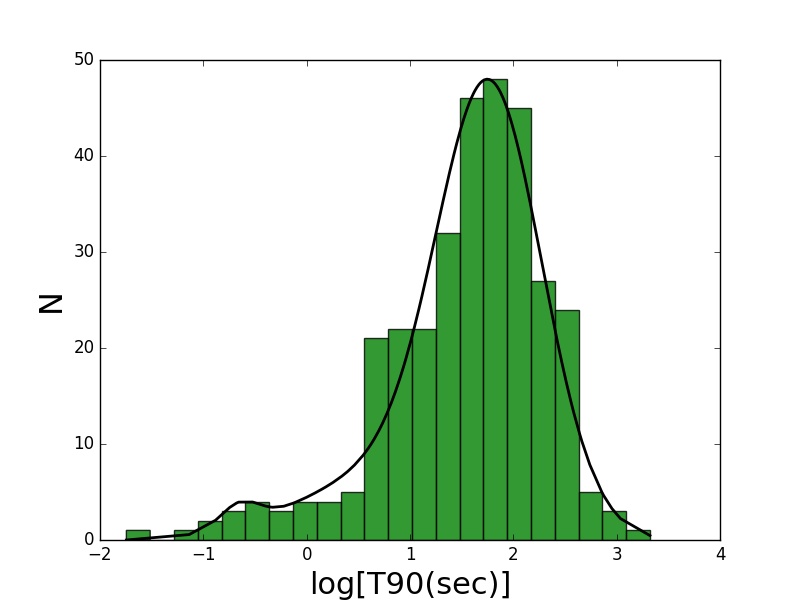}
\caption{A fit for the 3-component model for red shifted SWIFT GRBs with measured redshifts the observer frame. Summary of the fits can be found in Table~\ref{tab:redss_new}.}
\label{fig:redss_new2}
\end{figure}

We now carry out a similar classification of the durations of the GRBs in the observer frame for which we have the measured redshifts. The Swift GRB catalog consists of 323 detections with redshifts. The model fits for the intrinsic T90 for the SWIFT  GRBs are shown in Fig.~\ref{fig:redss_new1} and Fig.~\ref{fig:redss_new2}  respectively for the $k=2$ and $k=3$ case and a tabular summary of the model comparison results in Tab.~\ref{tab:redss_new}. On fitting the data to two components, we find that 64 and 259 GRBs\
 belong to the short and long category respectively. On doing the same for three components, 5, 69, and 249 fit in \
short, intermediate, and long categories respectively.


\begin{table*}[!htbp]
\caption{Model Comparison Parameters for the Swift GRBs which have a measured redshift value in the dataset. Explanation of all columns is same as in Table~\ref{tab:batse}. The preferred values are highlighted in bold. All tests prefer two components.  However the significance is marginal in all the cases.}
\label{tab:redss_new}
\begin{tabular}{|c|ccc|ccccc|ccc|}
\hline
$k$ & $ \mu_{T90}$ & $\sigma_{T90} $ & $w_{i}$ & $\mathcal{L}$ & $\chi^{2}/\nu$ &P$(\chi^{2},\nu)$   & $AIC$ & $BIC$ & $p$-value & $\Delta(AIC) $& $ \Delta(BIC)$ \\
\hline
\multirow{2}{*}{2} & 0.661 & 0.914 & 64 & \multirow{2}{*}{1522} & \multirow{2}{*}{\textbf{1.08}} & \multirow{2}{*}{\textbf{9.019e-3}} & \multirow{2}{*}{\textbf{697.1}} & \multirow{2}{*}{\textbf{712.2}} & \multirow{5}{*}{0.565($<0.02\sigma$)}&\multirow{5}{*}{2.1} & \multirow{5}{*}{9.7}\\
\cline{2-4}
& 1.766 & 0.5148 & 259 & & & & & & &  &\\
\cline{1-9}
\multirow{3}{*}{3} & -0.6472 & 0.189 & 5 & \multirow{3}{*}{1523} & \multirow{3}{*}{1.084} & \multirow{3}{*}{8.66e-3}   & \multirow{3}{*}{699.2} & \multirow{3}{*}{721.9} & & & \\
\cline{2-4}
& 0.8406 & 0.888 & 69 & & & & & & &  &\\
\cline{2-4}
& 1.776 & 0.507 & 249 & & & & & & &  &\\
\hline
\end{tabular}
\end{table*}

The observed T90 distribution for the Swift GRBs with recorded redshift values weakly prefers the $k=2$ case over $k=3$ case with all the three model-comparison tests. The frequentist model comparison test shows a $p$-value of $0.565$ which does not conclusively assert any preference similar to the inference from the $\Delta$AIC = 2.1 which only gives weak support to the $k=2$ model over the other. The $\Delta$BIC = 9.7 value however prefers the $k=2$ model with strong evidence, but not decisive enough. Therefore, it can be concluded that the GRBs with measured redshifts prefer the $k=2$ model with low to moderate confidence. Our results qualitatively agree with similar analysis done for recent SWIFT T90 distribution in the observer frame for GRBs with measured redshifts, by~\citet{Yang} and \citet{Tarnopolski}, both of whom find that two components are favored compared to three.

\begin{table*}[t]
\caption{Summary of model  comparison tests for all the different GRB datasets analyzed. The last two rows summarize the analysis of subset of SWIFT GRBs with measured redshifts in the  intrinsic frame and observer frame respectively. We find that only for the Swift GRBs (in the observer frame) three components are preferred with very decisive evidence using information-criterion-based tests and $2.36\sigma$ significance from frequentist model comparison tests.}
\label{tab:summ}
\begin{tabular}{|c|cc|cc|cc|}
\hline
\multirow{2}{*}{Dataset} & \multicolumn{2}{c}{$p$-value (from $\chi^2$ probability)} & \multicolumn{2}{c}{$\Delta AIC$} & \multicolumn{2}{c}{$\Delta BIC$} \\ 
\cline{2-7}
& Model preferred & Magnitude & Model preferred & Magnitude & Model preferred & Magnitude \\
\hline
BATSE & 3 & 0.109 & 3 & -6.5 & 2 & 5 \\
\hline
Fermi & 3 & 0.321 & 3 & -1.7 & 2 & 10 \\
\hline
BeppoSAX & 2 & 0.256 & 3 & -1.5 & 2 & 8 \\
\hline
Swift & 3 & 0.009 (2.36$\sigma$) & 3 & -28 & 3 & -18 \\
\hline
Intrinsic Swift & 2 & 0.518 & 2 & 1.7 & 2 & 9.8 \\
\hline
Swift (with measured $z$) & 2 & 0.565 & 2 & 2.1 & 2 & 9.7 \\
\hline 
\end{tabular}
\end{table*}

\section{Conclusions}

The main goal of this paper was to investigate the existence of an intermediate class of Gamma Ray Bursts,  in addition to the pre-existing short and long class of GRBs,  as previously argued by several authors. We did a comprehensive analysis of the T90 distributions of GRBs from all the major instruments used to detect them in the past three decades, by fitting the data to two as well as three lognormal distributions.  We then conducted three statistical tests to ascertain the best model among these two hypotheses. These tests include AIC, BIC, and a frequentist model comparison test based on  $\chi^2$ probability. The statistical significance from the information criterion based tests was obtained using empirical strength of evidence rules. From the frequentist test, significance was obtained by using Wilk's theorem.

 Our results for each of the detectors are as follows. A tabular summary of all these results can be found in Table~\ref{tab:summ}.
\begin{enumerate}
\item  For the BATSE dataset, we find that the frequentist model-comparison test and AIC prefer three components, whereas BIC prefers two. However, the significance from all the three tests is marginal and hence the evidence for the third component is weak.
\item  The results of model comparison tests for Fermi-GBM are same as that for BATSE. Both AIC and the frequentist model comparison test prefer three components, whereas BIC prefers two. However, the statistical significance from all these tests is quite weak and no decisive evidence can be made.
\item For BeppoSAX GRBs, AIC prefers three components, whereas BIC and frequentist model comparison tests prefer two.
However the statistical significance of each of these tests is marginal and no decisive evidence can be made either way.
\item For Swift GRBs, all three tests favor three components. The statistical significance of the third component
from the frequentist model comparison test is about 2.4$\sigma$.  Both $\Delta$AIC and  $\Delta$BIC value points to decisive evidence for a third component.    
\item Since a large number of  Swift GRBs have measured redshifts, we redid the classification on the intrinsic T90 distribution for the Swift GRBs. All the three tests favor two GRB components. The statistical significance though is very marginal.
\item For the subset of SWIFT GRBs with measured redshifts, we did the same classification in the observer frame and we find that all tests favor two components, although with marginal significance.
\end{enumerate}

Therefore in conclusion,  we find that none of the detectors show consistent results in accord with previous findings~\citep{Gehrels,Yang}. We should also point out that for most of  the datasets analyzed, not all the three model comparison methods agree with each other. However, we find that when there is a  disagreement, the significance from any one test  is marginal.  For the Swift GRB datasets in the observer frame, we find that   both the information criterion tests give consistent results  with $\Delta$IC$>10$ in favor of the three component model. For this dataset, the frequentist model comparison test is also consistent with  three components at about 2.4$\sigma$ level. Therefore all three tests agree in favor of the three-component model for Swift GRBs in the observer frame. However, when we carry out the same test with the intrinsic and observed T90 distribution for a subset of Swift GRBs with measured redshifts, we do not find evidence for the third component. Both these subsets show evidence for two components, although with not very high significance. For all the other detectors, the evidence for the third component is either very marginal or  disfavored.

\acknowledgments
We would like to thank Peter Veres and the anonymous referee for valuable feedback and comments on the paper draft.

\bibliographystyle{apj}
\bibliography{grb}

\end{document}